\documentclass{physeauth}
\usepackage{graphicx}
\usepackage{amsmath}
\usepackage{amssymb}
\begin{document}

\begin{frontmatter}

\title{Physics at the FMQT'04 conference}
\author[prag]{V. \v{S}pi\v{c}ka\thanksref{thank1}}
\author[amst]{Th.M.~ Nieuwenhuizen},
\author[det]{P.D.~Keefe}
\address[prag]{Institute of Physics, Academy of Sciences 
of the Czech Republic, Na Slovance 2, 182 21 Praha
8, Czech Republic}
\address[amst]{Institute for Theoretical Physics, 
Valckenierstraat 65, 1018 XE Amsterdam, The Netherlands}
\address[det]{Keefe and Associates, 24405 Gratiot Avenue, Eastpointe, Michigan 48021 USA}

\thanks[thank1]{
Corresponding author.
E-mail: spicka@fzu.cz}

\begin{abstract}
This paper summarizes the recent state of the art of the following topics 
presented at the FQMT'04 conference: Quantum, mesoscopic and (partly) 
classical thermodynamics; Quantum limits to the second law of thermodynamics; 
Quantum measurement; Quantum decoherence and dephasing; Mesoscopic and nano-electro-mechanical
systems; Classical molecular motors, ratchet systems and rectified
motion; Quantum Brownian motion and Quantum motors; Physics of quantum
computing; and Relevant experiments from the nanoscale to the macroscale.
To all these subjects an introduction is given and the
recent literature is broadly overviewed. The paper contains some 450
references in total.

\end{abstract}

\end{frontmatter}

\section*{Introduction}
The recent advancement of  technology has enabled very sensitive 
experiments on natural and artificially prepared systems of molecular sizes. 
The possibility to shape such experiments provides many challenges
from the point of view of understanding of basic concepts of physics 
related to these systems and development of methods for their 
description. There are two essential differences between these
"mesoscopic" systems we have in mind here, and large extended 
systems, such as crystals, described by the common thermodynamics and
statistical physics theory.  First of all, the typical "mesoscopic" 
system is of the intermediate size range between microscopic 
and macroscopic sizes. Second,  the system can consist of only  
a relatively small  amount of particles.  
The system is, however, very often connected via  
interactions with a macroscopic reservoir.
This is very different from the situation
which we use to describe by standard thermodynamics,
where both, systems and reservoir, are large extended
systems and the state of the system can be well 
characterized by macroscopic characteristics, 
like temperature. As for "mesoscopic" systems we 
definitely have to reconsider our concept related
to the description of the system. In addition,
due to their smallness, many of these "mesoscopic" systems
can manifest quantum behaviour. Manifestations of quantum features
like interference effects depend, of course,  on the characteristic
lengths of the system and temperature of the reservoir. 
Recent technology enables us to change very fine details of systems and 
conditions of measurements and to test various theoretical 
concepts experimentally. We are thus forced by  technology 
and experiments to  understand many essential 
concepts of the quantum  theory, thermodynamics  
and statistical physics in this new context. It is
not a trivial task at all to decide what characterizes such small
systems and what information we can gain from our
measurements. The characteristic phenomena for these
"mesoscopic" systems are quantum coherence and decoherence,
("thermal" and quantum) fluctuations and related noise 
in measured characteristics, tunnelling effects and 
dissipation. Under these conditions it is really
hard to create a theory for the behaviour of  the "mesoscopic" 
analogies of heat engines and motors, the themes opened
by  classical thermodynamics. Not surprising at all,
the question of the validity of various formulations of the 
Second Law of thermodynamics in such systems has emerged. 
Apart from this,  we are experimentally in touch not 
only with the basics of thermodynamics and 
statistical physics but also with
quantum theory itself,  since more and more precise
experiments on these  "mesoscopic" systems also challenge
the interpretation of the quantum theory, its completeness 
and related theory of measurement. Many models and 
experimental systems have their classical and quantum
version. Molecular motors and ratchets are considered
as classical or quantum systems depending on the 
parameters of these systems and their surroundings.
One of the main purposes of modelling and creating 
nano-electro-mechanical  systems (NEMS) is to study the quantum 
features of both electronic and mechanical parts of these systems, 
their interplay and to observe and better understand 
the transition between classical and quantum behaviour. 
The proper understanding of classical and quantum 
features of microscopic and macroscopic states and 
their relation to the decoherence, dephasing, relaxation
of systems, dissipation  and quantum measurement problems 
is needed to understand behaviour  of small "mesoscopic" systems. 
Since during measurements, systems can be very  far 
from equilibrium we have  to understand "arrow of time" problems, 
the emergence of non-equilibrium in these systems. To
develop  methods for the description of
"mesoscopic"  systems out of equilibrium and their
relaxation to equilibrium is the absolutely necessary
aim. It seems now that for the
proper, coherent, operational behaviour of "qubits systems" which could
lead to quantum computers in the future, the far from
equilibrium regime could be  the essential one.
At the same time,  solving the problem of
how to read-out the information from these
quantum qubits and not to disturb their coherence
essentially, a deep understanding of the relaxation
and dephasing processes is unavoidable.

Nowadays, all the above mentioned problems
connect thermodynamics, statistical physics, 
quantum theory and physics of small systems not only from a theoretical, 
but also from an experimental point of view, at many levels.
This recent state of the art motivated the organization of
the FQMT'04 conference and the following choice of its main topics:
Quantum, mesoscopic and (partly) classical thermodynamics;
Quantum limits to the Second Law of thermodynamics; Quantum measurement;
Quantum decoherence and dephasing; Mesoscopic and nano-electro-mechanical
systems; Classical molecular motors, ratchet systems and rectified
motion; Quantum Brownian motion and Quantum motors; Physics of quantum
computing; and Relevant experiments from the nanoscale to the macroscale.

Many participants have submitted a contribution to these proceedings. 
These have been grouped in five sections:\\ 
1. Quantum thermodynamics,\\
2. Quantum and classical statistical physics,\\
3. Quantum measurements, entanglement, coherence and dissipation,\\
4. Physics of small quantum systems, and \\
5. Molecular motors, rectified motion, physics of nanomechanical devices.\\

The grouping  has been made as much as possible on objective criteria
according to the prevailing orientation of the contributions.
Due to the complexity and often general aspects of solved problems 
and their overlaps with many areas of physics, most contributions could be, 
however, placed into at least two sections and the division
into sections is in the end, in some sense, a rather subjective
and artificial one providing only the first, very rough,
orientation between contributions.

\subsection*{A guide in the bibliography}
The details of the recent development regarding to the subjects
of individual sections (altogether with some very recent development
during a period of several months after the conference) can be found 
in the included literature (ordered mostly by years of publication): 

\noindent
{\bf 1. Quantum thermodynamics:
from \cite{BR86} to \cite{HMH05}.}
\\
{\bf 2. Quantum and classical statistical physics: 
from \cite{Hanggi82} to \cite {LEUWEN05}.}
\\
{\bf 3. Quantum measurements, entanglement, coherence and dissipation:
from \cite{EPR35} to \cite {MILTON01}.}
\\
{\bf 4. Physics of small quantum systems: 
from \cite{FEY66} to \cite {ASP05}.}
\\
{\bf 5. Molecular motors, rectified motion, 
    physics of nanomechanical devices:
from \cite{ASTUMIAN97} to \cite {BLEN05}.}
\\
We note that, apart from some exceptions, only recent books and review articles 
are referred to. We suppose that the reader will find all other important
articles in these books and reviews. Apart from this, we often do not
refer in the text to  specific books or review articles and leave
up to the reader to find out the more detailed information
from the variety of references offered in this article,
which are roughly classified above. To help the reader, 
all references are given with
their titles, not only books,  but also all articles.

\subsection*{Contents}
The aim of this article is to summarize the problems
discussed at the conference, to introduce main topics
of individual contributions and, last but not least, 
to point out relations between these topics. 

The following five sections of this article 
correspond to the five groups of the contributions
to these proceedings.

Each of these five sections  consists of two parts:
in the first part, the problem of the section is  introduced.
In the second part, called contributions to the conference,
a short summary of all contributions to the proceeding section
is given. Contributions are commented in the order in which
they are published in the proceedings. 

Due to many relations between discussed topics, texts
in the following five sections partly overlap. The aim
is, however, to show common themes from  different
points of view and levels of generality in different sections. 

\section{Quantum thermodynamics}

This was the subject that Vladislav \v C\'apek worked on in the last decade or 
so of his life (see reference in his book with Daniel 
Sheehan \cite{CAPEK05}) and  it was the original motivation 
for the conference. 
Its covering continues the line started in the conference
{\it Quantum limits to the Second Law} 
organized 
(Organizing committee: V.~\v C\' apek,  Th.M.~Nieuwenhuizen,  A.V.~Nikulov, and   D.P.~Sheehan)
at the University of San Diego (USA) 
in July 29-31, 2002 \cite{SANDIEGO02}, where Vladislav was a co-organizer, 
and was continued in the Lorentz workshop (organized by Th.M.~Nieuwenhuizen, M.~Grifoni,
and  E.~Paladino)   
{\it  Hot Topics in Quantum Statistical Physics: q-Thermodynamics,
 q-Decoherence and q-motors},  
that took place August 11-16, 2003, Leiden (the Netherlands).

Originally, thermodynamics developed as the phenomenological description
of the macroscopic behaviour of macroscopic systems. It formulated
the most general laws of the macroscopic world as the First and the Second Laws
of thermodynamics and introduced such concepts as temperature, heat, 
entropy and state variables. Phenomenological theory of heat engines 
based on thermodynamical behaviour of 
macroscopic systems was also developed. Later on, Boltzmann and his followers
created statistical thermodynamics. The concepts of micro-states and macro-states
of a system were created and dynamics of systems at the microscopic 
level were connected to the averaged, macroscopic, behaviour of the system.

When quantum mechanics appeared, statistical thermodynamics had to take
into account additional ingredients, but the overall structure of 
thermodynamics and its laws, and its meaning as the method of description 
of  huge,  macroscopic systems, remained unchanged since 
it was believed that quantum mechanics does not play a 
role at the macroscopic level. 

The real challenge for thermodynamics came
with the miniaturization of systems which were 
the objects of experiments. In addition, discussions about 
macroscopic quantum effects and 
possible interference of macroscopically distinct states
also contributed to a new emerging view of thermodynamics.
The question emerged under which conditions the thermodynamic behaviour 
still manifests. And, of course, whether the thermodynamic laws are 
still valid. Additional quantum mechanical ingredients as
quantum interference effects, (coherent) tunnelling, 
quantum non-locality and entanglement, quantum (not only thermal) fluctuations
and finite size systems (splitting to system and reservoir)
together with possible reduced dimensionality of systems, 
started to play an important role. All old certainties, as
the theory of heat engines, Maxwell's demon problem, its relation
to information and thermodynamics laws, appeared suddenly in a new 
light. Discussions about what is the meaning of quantum thermodynamics
started and  continue up till today, together with a huge development in the related field
of the quantum statistical and mesoscopic physics, see also Sections 3,4,5. 
The theoretical considerations have been complemented  by more 
and more sophisticated and  sensitive, sometimes really "crafty", experiments.  
In fact, there is the question up to which extent (size, parameters of systems) 
thermodynamics can provide unifying description of  "macroscopic objects" based
on the laws known from  statistical physics (discussed in the
next Section 2) and quantum mechanics (Section 3). 
Especially, the use of the concept of temperature and its limits 
were questioned in connection with small quantum systems. 
The validity of the Second Law of thermodynamics was questioned, too.
New suggestions of "heat" engines on the molecular
level have been discussed. 
In addition, concepts developed in these three inter-related disciplines
(discussed in Sections 1-3 of this article) are nowadays intensively 
tested and their possible limitations manifested 
by experiments on small quantum (mesoscopic) systems
(Section 4), which special cases as molecular motors and 
nano-electro-mechanical systems (NEMS) are discussed in Section 5.

\subsection{Contributions in the proceedings}
The main players in this field present their contributions. 
First, there is the Scully group (Texas A\&M; Princeton University) 
that focuses, in a series of papers on quantum optical engines, 
their fight against the Maxwell demon, and explain this old paradox
on the basis of quantum thermodynamics.

Next, there is the Mahler group (Stuttgart), which, together with
Gemmer (Osnabr\"uck), presents a
long argument for the emergence of thermodynamic behaviour in small quantum
systems by introducing random quantum states.
Closely related is the clarification of the question of when the notion of
temperature applies to small quantum systems.

In his opening talk of the session, Nieuwenhuizen started out 
from the First and Second Laws as they apply to finite and nanoscale systems, 
integrating it with the work of \v C\' apek.
He stressed that some formulations of the Second Law can be violated, 
though no case is known where they are all violated.
A new part of the material, referring to non-optimality 
of adiabatic work processes, is presented here; 
overviews of the further material of the talk are mentioned.

Then there are contributions of other long time players in the field: 
Ford and O'Connell discuss properties of the fine-grained entropy; 
Sheehan works out experimental setups which can be tested;  
Berger works out a description for the Chernogolovka experiment on power production
by inhomogeneous mesoscopic rings. 
Keefe describes how a conventional superconductor
may have an unexpected efficiency 
when cycling it across the transition line.

Patnaik and other members of the Scully group also clarify
the role of injection times in certain lasers without inversion.

\section{Quantum and classical statistical physics}

Statistical physics is the powerful approach to study macroscopic properties of systems
for which the dynamics is by far too difficult to study otherwise than numerically.
It has provided a theoretical basis of the laws of thermodynamics due to its recognition
of the molecular structure of matter, and has applications to a diversity of systems 
with many elements, also outside the range of condensed matter physics, 
such as star clusters, granular materials, traffic problems, 
econophysics, risk management, etc.

The basic task of  statistical physics is to relate microscopic
characteristics of the systems, like interactions and dynamics 
of their many microscopic parts, with their macroscopically
observed properties. It connects the level of description
of the dynamics of individual particles, such as electrons, 
with macroscopic behaviour of such complicated structure, such  
as metals. Special attention must be paid to the description
of the systems when the amount of particles involved is
somewhere between microscopic and macroscopic,
e.g. it has mesoscopic features, see also Section 4
and references there.  Statistical mechanics  is our tool to 
understand at least partly  (in general non-equilibrium) 
many particle interacting systems  and phenomena related to 
these systems as are various transient, relaxation, transport 
and dissipation processes, (thermal) fluctuations  and 
corresponding noise during  measurements on systems - in summary 
to understand all (generally non-linear and non-equilibrium)  
stochastic processes, linear or non-linear effects, short and long 
time behaviour of systems and dependency of the behaviour of 
an individual system on its initial state, structure, 
size and dimensionality.  This is accompanied by better
understanding of the reversibility of phenomena 
at the microscopic level  and the general irreversibility  
at the macroscopic level.  All the above discussion is common for both 
classical and quantum statistical physics. The properties of systems 
where quantum mechanics plays an important role, can be, however, 
in addition to classical behaviour, strongly influenced mainly
by the three essential manifestations of quantum mechanics:
the Pauli exclusion principle, quantum interference effects and 
quantum fluctuations. We will discuss in more detail 
mainly quantum interference and its relation to quantum
decoherence and dissipation in the next Section 3.
Quantum interference effects also play an essential role
in the mesoscopic structured discussed in the Section 4. 
In the following discussion we will take quantum mechanics 
into account.
 
Considering the huge variety of properties and phenomena 
related to various  systems to find the most feasible methods of 
their description, statistical physics has developed many methods.
Here we will mention only some which are the most relevant to
the conference contributions. 

One of the most important concepts for various systems descriptions
is the concept of closed and open systems.

\subsection {Closed Systems}
The theoretical microscopic description of any quantum system 
starts from the Hamiltonian of the isolated system which can be, 
however, driven by some external time dependent field described by the additional
time dependent part of the Hamiltonian. Such an isolated externally
driven system is then called a closed system.  
The dynamics of a closed system are governed
by the unitary evolution which is described either by the
Schr\" odinger equation for the wave function or the Liouville equation
for the density matrix of the system.
Very often the needed (relevant) observables are single particle
ones and, in this case, a one particle reduced density matrix
description is used to find these observables. 
This reduced one particle matrix is found from approximations
of the famous BBGKY chain of equations for reduced density
matrices \cite{LIB90,B98}.

From the point of view of formulation of an 
approximation scheme, it is  often advantageous 
not to calculate directly the single particle reduced
density matrix, but to formulate dynamics within the
Nonequilibrium Green's function method \cite{KB62}-\cite{ORR02}. 
This method was
extensively used for investigations of many extended
systems such as metals, semiconductors, plasma physics and nuclear matter
physics systems when the closed system description appears
as the natural one and, in consequence, it leads to
a solvable description of the system. 
Similarly to the closed equation for a single particle reduced 
density matrix obtained by approximations  within the BBGKY hierarchy,  
the irreversibility of the description and the
related description of the dissipation phenomena 
emerge in this description
when the asymptotical (approximal) equations
are closed either for the single particle Green's functions
or related single particle distribution function, as 
is in the case of the Boltzmann equation.

There are many identities and relations which help
to solve the dynamical equations written for 
closed systems. One special identity, which is worth 
mentioning here, is the famous fluctuation-dissipation 
theorem, which, as its name implies, relates fluctuations 
with the effect of dissipation. This theorem is at the heart of
linear response theory and enables us to formulate
Kubo-Greenwood formulas for solution of various
linear response problems. As an identity, which must
be fulfilled in any linear response theory, the 
fluctuation-dissipation theorem can also serve us
as the control for models involving dissipation.
The study of glasses has taught us, however, that it
applies only to systems where the largest timescale is 
less then the observation time \cite{BOUPHYS96,NIEUWPRL98}.

For better description of molecular, mesoscopic and quantum
optical systems,  it can be, however, advantageous 
(from the point of view of the possibility
to find the solution of the dynamics) to introduce 
the concept of an open system, which can be also
useful from the point of view of a more natural description of 
quantum mechanics itself, due to its principal non-locality,
see the following Section 3.

\subsection {Open systems}
Supposing there is a small part of a total system $T$, which
we are preferably interested in.  In this case, we divide 
the total closed system $T=S+B$, which is always governed by 
unitary evolution, to a so-called (relevant for us) 
open system $S$ and (irrelevant for us) a  bath $B$ named 
sometimes also a reservoir. The dynamics of the open system $S$ is 
then governed by the  non-unitary dynamics for the reduced 
density matrix of the  system $S$ obtained by projection of the total density 
matrix to the subspace $S$ and the Liouville equation 
for the total system $T$ only to the subspace $S$, too.
As a result of a projection technique, e.g. Nakajima-Zwanzig, 
we will have a  Generalized Master Equation (GME) for the reduced density
matrix of the open system $S$  \cite{Z64} - \cite{KF04b}. 
Formally,  this scheme
works pretty well. The first important problem, however,
emerges just at the level of this step. There are no essential
problems to find a reasonable approximation of the resulting
equations when the coupling between the open system $S$ and the 
bath $B$ is weak, so the separation seems to be quite natural.
In this case of weak coupling, we have the very well formulated
Davies theory. As soon as the coupling is very strong, problems
start and even today no really satisfactory approximations
are known. In this respect it is interesting to recall
the breakdown of the Landauer inequality for the amount of work
to be dispersed in order to erase one bit of information,
occurring exactly in this regime \cite{AN01,NA02}.

Generally, the GME has a very complicated structure, and
to find its solution for different systems and conditions 
is one of the tasks of recent quantum statistical
physics. Similar to the situation in the description
based on the closed systems, the basic approximation, which
essentially simplifies the GME, is the Markovian approximation
which removes all memory effects and introduces a local time structure
of the equation. In such a  case, the memory effects
are important for the description of the system, and much more
complicated non-markovian approximations are used.
From the point of view of behaviour of systems we can 
also formulate the GME in the so called Brownian motion
or Quantum optics limit - the names of approximations
and their use are self-explanatory.

Apart from methods based on the density matrix description
and related Liouville equation for the quantum mechanical
density matrix,  there are also methods using the path-integral 
formulation. Especially, the Feynman-Vernon formulation is 
often used. The Path integral formulation is especially
advantageous for formulation of problems with dissipation.
On the other hand, we can solve a dissipative quantum dynamical 
problem with a path integral approach, the GME,  or even 
via a generalized quantum Langevin equation.
Special attention to various models with dissipation
will be paid in the next section.

There are also recent attempts to combine the advantages
of Nonequilibrium Greens Functions (NGF), originally developed
within the concept of the closed systems, with the concept
of open systems by generalization the NGF method for
open systems \cite{KF04,KF04b}. 
This approach needs, however, still some time
to be developed into a practical working scheme.

\subsection{Contributions in the proceedings}
First, there is  a very elegant approach by Sk\'ala and Kapsa 
who derive the laws of quantum theory,  and the limit to
classical mechanics, on the basis of 
probability theory. 

The section continues with contributions from the Stuttgart/Osnabr\"uck groups
(represented by Gemmer, Mahler and Michel) on quantum heat transport 
and a relation between Schr\"odinger and statistical dynamics.
Next there is a contribution by a Prague group centered around Mare\v s on 
a classical problem put forward by the celebrated Prague scientist
F\"urth. They investigate a possibility to find the  difference
between classical and quantum Brownian motion in 
systems with periodic chemical reactions. 
The criterion for the experimental accessibility of 
F\" urth quantum diffusion limit is formulated in the article.
Experimental data show that the quantum nature of Brownian motion
in the investigated systems is very likely.

Men\v s\'{\i}k shows how to increase  chances
to solve complicated integro-differential 
dynamical structure of Nakajima-Zwanzig 
equations for the density matrix by 
transformation into the linear algebra system. 

In a series of three papers, \v Spi\v cka et al 
discuss  long and short time quantum dynamics
within the Nonequilibrium Green's function approach.
Reconstruction theorems for Green's function, which
enable construction of  single time quantum transport equations 
either of Landau-Boltzmann equation type for the
quasiparticle distribution function or General Master Equations
for the single particle density matrix, is 
discussed in detail.

After this there comes a contribution by 
Mare\v s et al on a method, called Stochastic Electrodynamics, 
that might underlie the well known but poorly understood zero point
fluctuations and zero point energy of quantum mechanics.

The article of de Haan deals with  
a resummation approach in classical physics that avoids infinities
such as the infinite self-energy of a point charge.

The follow up paper Khrennikov  attacks the claim that probabilities of quantum theory 
cannot be explained from classical probability theory; 
he explicitly shows that they follow directly, provided the
context (measurement setup) is specified first. 

Next there are contributions by Klotins on a
symplectic integration approach  in ferroelectrics and by Patriarca on the
Feynman-Vernon model for a moving thermal environment. 

The section ends with  a microcanonical approach to the 
foundations of thermodynamics by Gross.

\section{Quantum measurement, entanglement, coherence and dissipation}
This section deals with some core problems of recent physics,
as the foundations of quantum physics, mechanisms of
decoherence and dissipation and emergence of the classical 
world from the quantum one, as well as macroscopic 
irreversibility from  microscopic reversibility. These are,
nowadays, contrary to  past thinking, not only posed as theoretical, 
academic problems, but they are now more than in the past
reflected in  recent experiments and even suggested applications. 

The central phenomenon which connects such topics
as the quantum measurement problem, interpretation of quantum
mechanics, non-locality of quantum mechanics, quantum 
entanglement and teleportation, measurements on quantum systems
with possible quantum qubits behaviour and 
studies of various mesoscopic systems, is  the phenomenon 
of quantum interference. 

The existence of quantum interference, confirmed experimentally
at the microscopic level, brings the natural question about a possibility
of quantum interference of macroscopically distinct states.
This question is the basis of the famous Schr\" odinger's cat 
thought experiment \cite{ZUREK83}, which was formulated soon after 
another famous thought experiment, the Einstein-Podolsky-Rosen (EPR) 
paradox \cite{EPR35,ZUREK83}, questioning the completeness and non-locality of quantum 
mechanics. Both thought experiments ask 
the question what is the relation between the classical
and quantum physics. This leads  to other questions:
Where is the border line between the classical
and quantum worlds? What does macroscopic and microscopic
mean from this point of view? At which level can we still observe
superposition of quantum states? The standard Copenhagen
interpretation of quantum mechanics just states
that microscopic quantum objects are measured by 
classical macroscopic apparatus. The collapse
of the wave function (by some "stochastic" unknown process)
occurs in the relation with the measurement
and we will receive an "unpredictable" measured value. 
At the time of its formulation, experiments, which would enable
measurement of  the transition between the micro and macro worlds
under well defined conditions, were not accessible. 
With the possibility of more sophisticated quantum optics 
and solid state "mesoscopic" experiments, the old 
questions have re-emerged together with many new questions related
to the Copenhagen interpretation of quantum mechanics 
and other possible schemes for understanding the foundations of
quantum mechanics. Nowadays, however, these questions
can be discussed together with the relevant experimental
results.

The above mentioned problems were thoroughly discussed 
at the conference. Lively discussions about the foundations of quantum 
mechanics and related experiments followed talks
of leading experts in this area, R.~Balian, A.J.~Leggett and A.~Zeilinger.
They gave talks with the very fitting and self-explanatory
titles: \\ 
R.~Balian: "Solvable model of quantum measurement",\\
A.J.~Leggett: "Does the everyday world really obey
quantum mechanics?",\\
A.~ Zeilinger: "Exploring the boundary between  the quantum and
classical worlds".

Later on, K.~Schwab in his talk about Nano-electro-mechanical 
systems (see also Section 5),\\
K.~Schwab: "Quantum electro-mechanical devices: our
recent success to approach the uncertainty principle",\\
documented the very real and fruitful relations 
between fundamental questions of quantum physics, 
possibilities of the recent technologies and experimental 
physics dealing with small quantum systems (Section 4 and 5).

All these  lectures and the following discussions 
showed that the role of quantum interference and 
its erasing by decoherence processes is still
not fully understood, but we are gradually getting better
insights in many problems of quantum physics
of the micro-worlds and macro-worlds. In addition, we see
the old problems, represented by EPR and Schr\" odinger's
cat paradox in a new light. The emerging landscape of foundations of 
quantum physics and relevant experiments is more and
more complex. After pivotal experiments of Alain Aspects
and his group \cite{ASPECT76,ASPECT81,ASPECT82A,ASPECT82B,ASPECT99}
investigating  the non-locality
of quantum theory and Bell's inequalities 
from the late seventies and early eighties of the last century,
we have witnessed a wave of important experiments,
coming from two fields: quantum optics and 
solid state physics. 

Many experiments have attempted to test non-locality of quantum
mechanics as well as the quantum complementarity
principle. Since interference effects are often seen as the manifestation
of non-local behaviour, there is sometimes believed to be a direct
relationship between tests of quantum non-locality, entanglement
and complementarity. After Aspect's experiments (which tested directly
validity of Bell's inequality) other independent
experiments testing quantum non-locality appeared.
In 1989 Franson \cite{FRANSON89} 
suggested an experiment with energy-time entangled photons to compare
"standard" quantum mechanics with local hidden 
variable theories based on different degrees of interference
in these groups of theories. The corresponding experiments 
were realized about ten years later 
\cite{TITTEL98A,TITTEL98B,TITTEL99}.
These  experiments confirmed independently the
results of Aspect's group, i.e. strong violations of 
Bell's inequality. Another experimental scheme 
to test non-locality (using the idea of three-photon entanglement states, 
nowadays called GHZ states) was developed by Greenberger, Horne 
and Zeilinger \cite{GREEN90} and improved 
by Mermin \cite{MERMIN90}. The first experiments with GHZ states
were reported in 1999 \cite{BOUW99} and  quantum non-locality
was tested via three-photon GHZ states \cite{BOUW00} without
direct use of Bell's inequality. Recently, the question
of a single photon nonlocality has reappeared. For a recent
and "extreme" discussion for a single photon 
nonlocality, see \cite{MESSMOPRL04}; it is interesting  to compare
this paper  to the local interpretation by Vaidman \cite{VAIDMANPRL95}
and the related discussion \cite{HARDYPRL94,HARDYPRL95,GREENPRL95}.

The complementarity principle, which is in contradiction
with local theories, was tested via "which-way" double slit 
type experiments. A Gedanken which-way experiment
using micromaser cavities was suggested and
gradually improved upon by Englert, Rempe, Scully, 
and Walter \cite {REMPE91,SCULLY91,ENGLERT94,ENGLERT99}.
Ideas related to the so called quantum  eraser thought 
experiments  reported in the articles above were 
experimentally  realized in 1995 \cite{HERZOG95}. 
The quantum eraser principle was also lively 
discussed at the FQMT'04 conference after the lecture of 
Marlan Scully: "Quantum Controversy: From Maxwell's Demon
and Quantum Eraser to Black Hole Radiation".
 
All experimental tests of non-locality and complementarity
up to now support non-locality of the quantum mechanical
picture and seem to exclude the idea of local reality.
This is still a heavily debated subject, however,
and there are opposing view points as well, that 
argue that locality cannot be excluded, see e.g. 
\cite{KHREN01,KHREN01ACARDI,KHREN01MUYNCK,KHREN01VOLOV,HESS04}.
Non-locality is also strongly advocated on the basis of 
teleportation experiments using the entangled
states.  For the first time the possibility to teleport
a photon was discussed in \cite{BENNETT93}. Teleportation
was then experimentally realized in 1997 \cite{BOUW97}.

Another group of experiments related strongly
to both foundations of quantum physics, and even possible
applications, are experiments  dealing with 
the physics of quantum computing, i.e. physics of qubits. 
Several leading experts in this field 
delivered  their lectures at the conference speaking about
various aspects of the  physics involved, both from the theoretical
and experimental point of views. Namely, participants heard
(in addition to the contributions included in these proceedings)
the following lectures:
   
\noindent
B. Altshuler: "Non-Gaussian low-frequency noise as a source
of decoherence of qubits",\\
T.~Brandes: "Shot noise spectrum of open dissipative quantum
two level systems",\\
A. Caldeira: "Dissipative dynamics of spins in 
quantum dots",\\
H.~Mooij: "Coherence and decoherence in superconducting
flux qubits",\\
G.~Sch\" on: "Dephasing at symmetry points",\\
U.~Weiss: "Nonequilibrium quantum transport, noise and decoherence:
quantum impurity systems and qubits".\\

Again, as we can see even from the titles of these lectures,
the central theme of "qubits physics" is the 
theoretical description and measurement of 
three closely related phenomena: dissipation, noise and
decoherence. 

There are nowadays several ideas being put forward as how to
realize  quantum qubit systems practically.
The most active work is
mainly on these systems: 
quantum optical systems (and cavity quantum electrodynamics
based on systems), ion traps, liquid state
NMR, and spin systems in semiconductors. 
During the conference  special attention
was paid to superconducting circuit systems
which use the Josephson junction effect.
The common central theme of all the investigations
into these various systems is the fight between
quantum coherence (needed for the proper
function of qubit systems from the point of view
of possible quantum computing algorithms)
and decoherence (coming naturally from the environment
and being a natural obstacle to a realization
of possible "quantum processors" in the future, 
but is inevitable due to coupling 
to an environment  which enables us 
to read out information from systems).

In general, decoherence is a process of a loss
of quantum interference (coherence) due to non-unitary dynamics
of the system, which is a consequence of a coupling between the
system and the environment (in terms of theory
of open systems discussed in Section 2, due to
interaction between the open system $S$ and the 
reservoir $B$). Since technically quantum interference is
described by the off-diagonal elements of the
density matrix of the system, correspondingly 
the decay of these elements (their possible time 
development towards their zero  values limit)
describes the decoherence processes. When all
off-diagonal density matrix elements are zero,
the system is in a fully decoherent (classical) 
state. Phenomenologically, the transition in time
from the quantum (coherent) state into the classical
(decoherent) state can be described by 
a decoherence factor $e^{-t/\tau}$,
where $\tau$ is the decoherence time. 
Generally, the decoherence, of course,
includes both dephasing and dissipative
contributions (and not only), sometimes denoted as $T_2$ and $T_1$
processes. Dephasing is related to processes randomizing
the relative phases of the quantum states. Dissipation
corresponds to interaction processes which
are changing the populations of quantum states. 

The description of the decoherence processes 
for various systems is  a highly non-trivial task 
which is far from being satisfactorily fulfilled.
Many highly successful models have already been
introduced for the description of systems with dissipation,
e.g. variants of the central spin model (both, system
and reservoir are represented by spins), spin-boson
model (system composed by spins, reservoir by bosons)
not to mention the celebrated Caldeira-Leggett model.
However, as the conference talks and discussions revealed,
new, more complex and more realistic models are needed
to describe the dissipation processes together 
with  improvement of the general theory
of open systems, see also Section 2. There are still
many unanswered questions related to quantum
coherence, the most important, at least as it seems now,
are the following ones: \\

{\bf 1. What is the dynamics of decoherence?}
In other words, how do the off-diagonal elements of
the density matrix of the system evolve in time
under various conditions, depending e.g. 
on the initial state of the system and the reservoir, on the
strengths of coupling between the system and the reservoir? 
The realistic determination of decoherence times for
various systems is a very useful, but sometimes
difficult to fulfill, aim.

{\bf 2. What are possible mechanisms of decoherence
in various systems}?
Apart from this, what is the relation of these 
mechanisms  to other mechanisms in systems, e.g. namely
to quantum relaxation processes?
 
{\bf 3. What is the relation of decoherence
processes with the transition between 
the quantum and classical behaviour?}

{\bf 4. How are decoherence processes
related to quantum measurement process?}
Namely, a natural question emerges as to whether
the decoherence can cause collapse
of the wave function in relation to the
measurement processes. If yes, what is the difference
between measurement on microscopic 
and possible macroscopic coherent states, if any?
What is the relation to the possible irreversibility
on the microscopic level caused by quantum measurement?
In other words, can quantum decoherence satisfactorily solve
the "measurement problem" and related collapse
of the wave function, if this really occurs?\\

Investigation of various manifestations of quantum interference,
dissipation, dephasing and decoherence in general is
a very active area of recent research, since 
we need to understand decoherence at microscopic, mesoscopic 
and macroscopic scales to be able to deal with recent experimental 
systems, see also  Sections 4 and 5.
On the other hand, nowadays a huge diversity of investigated systems,
with often well controlled parameters, provide us
an enormous amount of experimental data to build
up a gradually more and more satisfactory picture
of decoherence processes and related theories of
their description. Apart from providing
a practical solution for every-day problems
encountered when analysing behaviour of 
experimentally tested systems, this progress
in knowledge about interference effects and
decoherence processes also helps us to improve our 
understanding of  quantum physics at its most
fundamental level. As already partially discussed above,
interference and decoherence play
a crucial role in interpretation of quantum 
mechanics and possible alternative theories.

Apart from the Copenhagen interpretation of
quantum mechanics and its small variations,
there are many other interpretations between
which it is difficult to distinguish since
they provide, at least in principle, the same
description of nature and the same results 
when applied to concrete physical situations.
Here we name only some important representatives
of these alternatives of the Copenhagen
interpretation:\\
1. Statistical interpretation as it is
represented by the approach of Ballentine
\cite{BAL70}, and embraced by Balian
on the basis of his solution of the quantum
measurement problem \cite{ABM01,ABNEURO03},\\
2. The de Broglie-Bohm interpretation with
de Broglie's idea of pilot waves and 
Bohm's idea of quantum potentials
\cite{BHK87,BOHM93,HP94,BH89},\\
3. Many world interpretation as represented
by Everett's approach \cite{EV73,ZUREK83},\\
4. Macroscopic realism as represented by Leggett's
contributions
\cite{LEGGETT80,LEGGETT86,DAV93,HP94,SAVITT95,LEGGETT02,EDK05},\\
5. GRWP (Spontaneous collapse models) theory as represented by the works
of Ghiradi, Rimini, Weber and Pearle
\cite{GHIRARDI86,GHIRARDI90,COLLETT03,BASSI03},
and by the recent development
in this field \cite{PEARLEQ05}, and\\
6. Penrose's theory combining quantum mechanics
   with the geometry of space and time
\cite{PENROSE89}.\\

We will not discuss these theories in detail here,
see many references to this Section at the end
of this article. We will just briefly comment 
that the problem of the collapse
of the wave function, measurement of 
microscopic versus macroscopic states and 
decoherence processes, are related in some
of the above mentioned interpretations
of quantum mechanics. Environmentally induced
decoherence is one of possible explanations of
the collapse of the wave function and non-possibility
to observe macroscopic superposition of states.

Generally, decoherence can be a candidate for explaining
most of the difference between the microscopic world of quantum
physics and the macroscopic (classical) world we  directly observe.
From this point of view, the idea of decoherence can 
help us in the end to understand, even at the very
fundamental level, the relation between 
quantum statistical physics and thermodynamics.
Since the decoherence time is very
sensitive to the parameters of the system
and to the reservoir with which the system is coupled,
its values can change over many orders from the
very small (non-measurable nowadays) values for 
macroscopic objects to the very large values for
almost isolated elementary particles.
The small, "mesoscopic" systems, see also Section 4,  however, provide
a possibility to make measurements of decoherence
in the time range which is observable by  recent
techniques.

\subsection{Contributions in the proceedings}
First, Balian et al contribute to the "perennial", but still
not satisfactorily closed,  discussion of the measurement of quantum
systems. The quantum measurement problem has long suffered from a lack of models with enough 
relevant physics, which has led to desperate views as being unsolvable, being a 
matter of philosophy, and so on. In his talk, Balian presented a simple,
yet sufficiently rich  model
for the measurement of a spin-$\half$. Based on the macroscopic size of the
apparatus, he connects the irreversibility of the measurement with 
the general problem of irreversibility in statistical physics, 
where the paradox of microscopic reversibility plays no role in practice, 
because it relates to unrealistically long times.
Balian also touched upon
questions related to decoherence (see points {\bf 1-4} above).
In the model he considered, the  Schr\" odinger cat terms 
vanish by dephasing and are, being hidden but still present, 
in a subsequent step erased by decoherence (the situation is similar 
to spin-echo setups, when no echo is made).
The registration of the measurement 
takes place on a still longer timescale and has classical
features. The whole setup, before, during and after the measurement,
has a natural look within the statistical interpretation of quantum mechanics.

One of the surprising features of quantum mechanics 
is its coherence and entanglement.
This leads to processes that, even though possible 
in the physics of classical Brownian motion, are rather unexpected. 
This theme is represented by works of 
B\"uttiker and  Jordan	on ground state entanglement energetics, 
of  Aharony, Entin-Wohlman and Imry	 on
phase measurements in  Aharonov-Bohm interferometers
of  Schulman and Gaveau	 on quantum coherence in Carnot engines, 
of  D'Arrigo et al. on
quantum control in Josephson qubits, 
and by Cohen on quantum pumping and dissipation.

\section{Physics of small quantum systems}
In the context of this section, 
systems are understood to be small (often also called  mesoscopic) 
when their parameters enable us to observe quantum 
interference effects manifested, for instance, in the transport characteristics
of electrons. Usually these systems are artificially created
structures which combine metal, semiconductor or superconductor
materials 
\cite{MAZ86,ALT91,DAV92,AKKER95,CER95,DATTA95,FERRY97,IMRY97,SOHN97,DAVIES98,SHIK98,SUP98,JAN01,MUR01}. 
Various characteristics related to electrons 
in these structures are studied. The "small" size of the 
system is not the only one decisive parameter which determines 
whether quantum interference will be manifested.
In fact, what is small from the point of view of 
manifestations of  quantum interference effects 
depends also on the interactions in the systems.
For instance, the quantum coherence of an electron 
which moves in the sample ballistically without scattering events 
can be disturbed by its scattering with phonons;
of course, with decreasing sample size there is
a bigger probability that the electron flows
through the sample without any inelastic
scattering which disturbs its quantum coherence.
On the other hand the increasing temperature
drastically increases the probability of 
electron-phonon scattering. So, when temperature
is lower, the size of the sample can be bigger
to observe interference effects related
to the electron moving without scattering through the sample. Of
course, the concentration of electrons is an
other parameter which influence the quantum
behaviour because of its relation with the
electron-electron interaction.

Physics of "small" (mesoscopic) systems has been
a very active area of research already for
many years, which brings further and further
motivation for investigations 
due to  ever improving technologies
\cite{BAL03,GOSER04,HEISS05,DOTSPHYSICA05,CUN05,CFN05,SEM05,ASP05,DREXLER92}. 
These enable the preparation of 
more and more interesting  samples with really 
well defined parameters 
and to measure more and more,  in the past 
inaccessible, details. Nowadays, experiments
can measure quantum interference effects
in a system and their dependence on various parameters
as for example: dimensionality of the sample
(quantum dots, quantum wires and various
two dimensional systems are common),
size of the sample  and its geometry, concentration of 
impurities (the number of scattering events can be varied), 
concentration of electrons, temperature of the sample
and its environment, and strengths of 
electric and magnetic fields.  

These artificially prepared systems 
enable us to test various hypotheses, methods and
theories developed in the above discussed
areas of  Quantum thermodynamics (Section 1),
Statistical physics (Section 2) and Physics
of quantum measurement, entanglement, coherence
and dissipation (Section 3).

In these small systems, many quantum interference
and fluctuation phenomena are studied 
under various conditions, among others,
weak electron localization,  universal conductance fluctuations, 
persistent currents, and tunnelling (resonant tunnelling). 
Special attention is also paid to  the Aharonov-Bohm effect, 
quantum Hall effects, and  quantum chaos
\cite{DHI98,ALT91,DATTA95,FERRY97,IMRY97,JAN01,MUR01,IHN}.

An especially fast developing area is  
"quantum dots" physics
\cite{NAK04,HEISS05,DOTSPHYSICA05}. Nowadays, quantum dots can be fabricated 
with a few levels, thus constituting artificial atoms. 
As their parameters can be manipulated,  this yields unprecedented 
tools to study the dynamics of few level open systems
and dissipative processes in a controlled way.
Quantum dots systems, as mentioned already in Section 3, 
are also candidates for creating working qubit systems.

Another very active area of research is dealing with 
molecular systems and molecular electronics
\cite{GOSER04,SHEEHAN05,CUN05}.

"Mesoscopic" systems also contributed to the development
of some special theoretical methods of  quantum
statistical physics. To describe very effectively
linear transport of electrons in mesoscopic systems, the
Landauer-B\" uttiker method was introduced
\cite{DHI98,MAZ86,DATTA95,FERRY97,IMRY97,IMRYRMP99}. This formalism,
based on the idea of transport as a scattering
problem, is  suitable for the description
of transport through samples where only
elastic scattering (on impurities) takes place. 
Transport channels are then well
described by transmission and reflection
coefficients, and we have a simple recipe of
how to calculate transport characteristics.
In this case, this efficient method is
equivalent to the Kubo-Greenwood formula
which has to be, however, used when inelastic
scatterings must be taken into account
\cite{VSM89,SMV90,MEIR92}.
To describe various transport regimes
in the case of disordered systems, 
random matrix theory \cite{BEEN99} and  non-linear sigma 
models \cite{EFETOV97} are also in use. Many techniques, originally used 
for the description of bulk (extended)
systems, as for example  Green's functions 
\cite{LAKE92,WINGR94,JAUHOWIN94,MICHAEL03,VSM89,MEIR92,DATS00,XUE02,W02,ZENG03,J03}
or the path integral
approach \cite{DHI98},  have been also adapted to 
describe the physics of small  systems.

\subsection{Contributions in the proceedings}
This section contains papers by Hohenester and Stadler	
on quantum control of the electron-phonon scattering 
in artificial atoms, of  Kuzmenko, Kikoin, and Avishai	
on symmetries of the Kondo effect in triangular quantum dots,
by Rotter et al. on Fano resonances and decoherence in transport 
through quantum dots, by  Kr\'al  and Zde\v nek on the
stationary-state electronic distribution in quantum dots.

Further there is a contribution by Kamenetskii on
mesoscopic quantum effects of symmetry breaking for 
magnetic-dipolar oscillating modes,
of Sharov and Zaikin on parity effects and spontaneous 
currents in superconducting nanorings and by  Sadgrove et al. on
noise on the quantum and diffusion 
resonances of an optics kicked atomic rotor.

The section ends with a  contribution from  Mare\v s et al which deals
with the weak localization from point of view of
stochastic electrodynamics.

\section{Molecular motors, rectified motion, physics of nanomechanical devices}
Physics of molecular motors and nano-mechanical
systems create special branches of physics of small
("mesoscopic") systems. Contrary to the preceding section, this section
deals with classical as well as quantum systems.

Contributions in the proceedings deal with many aspects
of molecular motors and rectified motion (classical and quantum versions
of ratchet systems)  and discuss various aspects of molecular 
and nanomechanical devices.

\subsection{Molecular motors and rectified motion}
The basic feature of ratchet systems is the existence
of a periodic, but asymmetric potential in the presence of
an ac driving field. In addition, a system with a ratchet
effect must have such parameters that thermal and quantum 
(in the case of quantum motors)  fluctuations  
play an important role in its dynamics. Under these conditions 
directed transport can appear both  in classical and quantum systems.
Due to the essential role that Brownian motion
plays in the ratchet effect,  systems manifesting
this effect are called either ratchet, or equivalently,
Brownian motor systems.  Due to the importance of fluctuations, 
ratchet effects appear generally in small systems
\cite{ASTUMIAN02,LINKE02,RH02,REIMANN02,HMN05}. 

The ratchet effect occurs naturally in biological systems
where it creates a base for functioning of  so-called molecular motors. 
These are proteins that take  care of transport and muscle 
contraction in living organisms
\cite{BIER97,SITGES99,BIO2000,HOWARD01,SCHLIWA03,HOW03,BIER05,FREYK05}. 
Apart from these
naturally created systems, molecular motors are also
studied in artificially shaped systems which, in some sense,
mimic functions of molecular motors in living
cells \cite{LINKE02,HMN05}.

There were several very interesting lectures at
the FQMT'04 conference which covered Brownian
motion and  molecular motors in both classical and quantum
variants together with relevant experiments and possible
applications. This can be demonstrated by the following
lectures presented at the conference:
H.~Grabert: "Quantum brownian motion with large
friction",\\
M.~Grifoni: "Duality transformation for quantum ratchets",\\
P.~H\" anggi: "Brownian motors",\\
H.~Linke: "Nanomachines: from biology to quantum heat engines",\\
T.~Seideman: "Current-driven dynamics in molecular scale
electronics. From surface nanochemistry to new forms of 
molecular machines",\\
S.~Klumpp: " Movements of molecular motors:
Random walks and traffic phenomena".\\
Klumpp  discussed a less studied aspect of these motors,
namely their large scale motion, which of course is the thing
that makes them so relevant in our bodies.

The theoretical and experimental study of both classical and
quantum molecular motors enables us to develop a better stochastic 
method of systems description which is in some sense complementary
to a fully microscopic description, starting from   
deterministic Newton or Schr\" odinger equations. Similarly, 
as Langevin and Fokker-Planck equations are complementary
to the reversible, deterministic Newton equation and irreversible
statistical mechanics based on it, the quantum Langevin equation 
and other quantum stochastic  equations are complementary to the 
irreversible quantum statistical description
starting from the "reversible, deterministic" Schr\" odinger equation. 
In the end, both approaches, either the one starting from the 
deterministic description or the one starting from the stochastic 
description, must provide the same results.
Again, natural questions 
in relation with classical
and quantum molecular motors, 
are "How the irreversibility is 
emerging?" and "Where is the crossover between classical
and quantum worlds?".

\subsection{Nano-mechanical systems}
In this subsection, we will briefly comment on two categories of small mechanical systems,
opto-mechanical and nano-electro-mechanical systems (NEMS). 

The central part of both systems is the mechanical resonator of nanometer to micrometer
size scale which is coupled to a specially shaped "environment". 
This coupling enables us to detect vibrational modes of the resonator 
and also enables these systems to work as "devices". 

Due to  advances in  microfabrication techniques, nanomechanical devices 
have a great potential, not only in applications, as e.g. 
ultrasensitive mass and force  detectors at the molecular level, 
high-speed optical signal processing devices, and electrometers, 
(e.g. when coupled to  a Cooper-pair box) but also in investigations
of fundamental concepts of quantum mechanics. 

{\bf Opto-mechanical systems} consist of a resonator coupled to a radiation field
by radiation pressure effects. A radiation field serves as a probe to read out
information about the state of the resonator (oscillator's frequency and position).

At the FQMT'04 conference, the lecture of  P.~Tombesi: "Macroscopic entanglement 
for high precision measurements" was devoted to  applications of opto-mechanical 
systems of high precision measurements \cite{TOMBESI01,TOMBESI04}.

In the following brief summary of nano-mechanical systems we will concentrate 
on the discussion of NEMS since they  were, in comparison to opto-mechanical 
systems, far more discussed at the conference. 
In addition, contrary to opto-mechanical devices, contributions 
related to NEMS are presented in these proceedings.

The overview of recent developments in the physics of NEMS was contained
in the following three lectures:\\
M.~Blencowe: "Semiclassical Dynamics of Nanoelectromechanical
systems"\\
A.~MacKinnon: "Theory of some nanoelectromechanical systems"\\
K.~Schwab:"Quantum electro-mechanical devices: our
recent success to approach the uncertainty principle"\\

{\bf Nano-electro-mechanical systems (NEMS)} are nanometer to micrometer
scale mechanical resonators coupled electrostatically to electronic (mesoscopic) 
devices of comparable size. In other words, NEMS are micro-electro-mechanical
(MEMS) systems scaled to submicron size. As a  central part of NEMS, the
mechanical resonator, very simple structures, such as a cantilever or a bridge, are commonly
used \cite{ROUKES03A,BLENCOWE04,BLEN05}.
A mechanical resonator having submicron size and small mass can
vibrate at frequencies from a few megahertz up to around a gigahertz.
There is a possibility to detect the displacement of the
vibrating part of the resonator (e.g. cantilever) by ultrasensitive
displacement detectors. Several working schemes
have been suggested \cite{ROUKES03B,BLENCOWE04,ROUKES05A}. One of the possibilities 
for the extremely sensitive motion detectors for the 
nanomechanical resonator is a single-electron  transistor 
(SET) \cite{KNOBEL03,ARMOUR04,ARMOUR04b}.

There are plenty of suggested, and even experimentally 
realized schemes for devices 
using electro-mechanical coupling to the submicron resonator.

{\bf Mass spectrometer:}
When a small particle (molecule) attaches itself to a
resonator, its mass can be determined from the resulting
vibrational frequency shift of the resonator 
\cite{EKINCI04,EKINCI04b}.

{\bf Electro-mechanical which-way interferometer:}
The resonator (cantilever) is electrostatically coupled to a quantum dot
situated in one of two arms of an Aharonov-Bohm
ring. The vibrating cantilever decides which way
the individual electron goes from the dot. At very low temperature
a submicron cantilever can be represented by a single 
quantum mechanical oscillator \cite{ARMOUR01,MACKINNON03}.

{\bf Quantum shuttle:}
This is a model device suggested originally by Gorelik 
et al in \cite{GORELIK98}. In this  model, a movable dot, 
coupled to a quantum harmonic oscillator, is situated
between two contacts. Electrons are shuttled from one
contact to the other on the dot.  In the vicinity
of the contacts, the electron can tunnel from the dot
to the respective contact. There are variants of this model
(single or triple dot arrangements). In addition, not
only shuttling of electrons, but also of Cooper pairs
has been studied \cite{SHEKTER03}. All the suggested models have
been intensively investigated from the point of view
of what are the proper observables which can decide between
quantum and classical shuttling processes
\cite{ARMOUR2002,NOVOTNY03,NOV04,NOVOTNY04}.
Recently, even full  counting statistics of the quantum shuttle 
model have been  calculated \cite{FLINDT05}.  
Even though there have been recent attempts to make 
quantum shuttles, it seems now that these devices
are still too large to be able to manifest
quantum effects.

{\bf Systems for solid state quantum information processors:}
There is a chance that nano-electro-mechanical systems will
play  an important role in the development of quantum
computer systems, see also Section 3. The task of fabricating
physical qubit elements in such a network that will
reach sufficiently long quantum decoherence decay times
and at the same time  will be able to control
entanglement of individual elements, is one of obstacles on our 
way to a quantum computer. Recently, a promising scheme
has been suggested: high frequency nanomechanical resonators could
be used to coherently couple two or more
current-biased Josephson junction devices to make
a solid state quantum information processing
architecture \cite{CLEL04,CLEL05}.

{\bf Nanomechanical resonators coupled to a Cooper-pair box:}
The system of a nanomechanical 
resonator which is electrostatically coupled 
to a Cooper-pair box has been studied both
theoretically and experimentally  
\cite{ARM02,IRISH03,BLENCOWE04,BLEN05}.
There has been a hope that these systems
can be used to test some ideas from the decoherence
theory and questions related to the foundation
of quantum physics, see the text below. 

{\bf BioNEMS:}
With advancing technologies and huge sensitivity
of NEMS to detect small inertial masses (even of individual molecules)
and at the same time forces (chemical forces), 
there is an increasing chance that NEMS will
be effectively used to improve  our knowledge of
macromolecules existing in living cells by measuring
their masses and binding forces.  Questions of the type:
"Can one realize a nanoscale assay for a single cell?"
have already been seriously asked. Biochips involving
nanoscale mechanical systems could be quite
helpful in biochemistry studies \cite{ROUKES00}. \\

Nano-electro-mechanical  systems represent a great hope 
for improving  our understanding
of many aspects of the behaviour of small systems.
Apart from providing ultra-sensitive measuring
techniques and many other possible applications, 
this also enables us to test basic ideas of
quantum statistical physics and conceptual
foundations of quantum mechanics mentioned 
in Sections 2 and 3.

\subsubsection{NEMS, statistical physics and foundations of quantum mechanics}
Taking into account "mesoscopic" sizes, masses of both the nanomechanical resonator 
and coupled devices, temperatures involved (NEMS systems operate
at  very low  temperatures) and in addition coupling of the whole 
NEMS into its surroundings, we can see that we have the systems {\it par excellence}
to study all essential questions of the quantum statistical
physics of open systems: fluctuations, noise, dissipation and decoherence
effects. For example, the analysis of the current noise spectrum 
can help to distinguish between possible mechanisms of 
transport of electrons between two contacts of a quantum 
shuttle device. Suggested models and approximation schemes 
can be tested experimentally.

Nano-electro-mechanical systems also offer a possible fascinating insight
into the realm of the foundations of quantum physics, since
their parameters approach now a possibility to
measure not only the crossover between classical and
quantum behaviour of a nanomechanical resonator, but
also to observe interference of macroscopically
distinct quantum states and related decoherence times,  due to
environmentally induced decoherence. In addition, NEMS are 
promising from the point of view of detailed
studies of  decoherence theory and of observations
of decoherence times which are important not
only for the tuning of NEMS and e.g. their possible
use for quantum processor systems, but also
for testing  alternative approaches to quantum
mechanics, where the decoherence times play
an essential role, see also Section 3.

{\bf A possibility to use  NEMS for which-way experiments},
one of the essential tests of interference behaviour and 
non-locality nature of  quantum mechanics, was already mentioned above.

{\bf Testing the Heisenberg Uncertainty Principle} is another choice.
There is an increasing effort to approach the quantum limit for
position detection. The recent \cite{HAYE04}
ultra-sensitive measurements of positions
of a resonator (effectively represented by an oscillator)
at very low temperature  were made on the NEMS system.
The positions of a nanomechanical  resonator, 
a vibrating mechanical beam (with the frequency of about 20MHz) 
which was about a hundredth of a millimeter long and cooled
down to about 60mK,   were measured by a 
single-electron transistor coupled  electrostatically  
to the resonator. It is fascinating to realize that
this test of the Uncertainty Principle used
a mechanical beam, very small from the point of 
view of human senses, but still macroscopic from
the point of view of common conception of microworld and
macroworld of quantum mechanics. The beam consists
of about $10^{12}$ atoms. Such a many-particle 
object definitely is not considered to be microscopic. 
This experiment is not only
trying to approach the Heisenberg Uncertainty limit
for a position measurement, but it tries to approach it 
for a macroscopic object. In other words,
this type of experiment aims to find a crossover 
not only between the quantum and classical worlds 
but also to find out how this crossover 
is related to the possible distinction between
the microworld and macroworld.

{\bf Interference of macroscopically distinct states and measurement of decoherence times:}
At the end of the discussion of NEMS 
and foundations of quantum mechanics, we will return
to the nanomechanical resonator coupled to Cooper-pair
box  NEMS already introduced above.
This NEMS offers a working scheme to produce superpositions of distinct position states and 
measure their decay due to environmentally induced decoherence
\cite{ARM02,IRISH03,BLENCOWE04}.
This scheme is based on the idea of coupling a nano-mechanical resonator
to a Cooper-pair box to gain an advantage of coupling the resonator to a well
defined two-level system (spin down and spin up states; a Cooper-pair
box consists of a small superconducting island which is linked through
a Josephson junction to a superconducting reservoir).
The aim is to produce entangled states of a mechanical resonator
and a Cooper-pair box: As soon as the Cooper-pair box is in a linear 
superposition of charge states (prepared by using an external
gate) the resonator is (due to entanglement) driven in a superposition
of spatially separated states. Under some circumstances, the separation
of these states is large enough so these states can be described
as distinct states. Since the used resonator (cantilever) contains
about $10^{10}-10^{11}$ atoms, we can suppose these states are
macroscopically distinct states. There is a possibility to observe
decoherence times related to this superposition of macroscopically 
distinct position states due to their coupling to the "well defined" 
environment.

\subsubsection{A guide in the bibliography}

The recent development in nano-electromechanical studies is well documented
in the book from A.N. Cleland \cite{CLELAND02} 
and several review articles \cite{ROUKES00,BLICK02,SHEKTER03,BLENCOWE04,BLEN05}.

\subsection{Contributions in the proceedings}
First of all, large scale motion of molecular motors is reviewed by Klumpp et al.
They use lattice models to deal with well-known traffic problems, in their case
in the context of motion of unbound molecular motors. In this way 
they model behaviour of molecular motors in living cells which are
responsible for driving the transport in organelles. 

Quantum heat engines based on 
particle-exchange are discussed by Humphrey and Linke.
They thoroughly discuss properties and differences in the thermodynamics
underlying the three-level amplifier (a quantum engine 
based on a thermally pumped laser) and two-level quantum heat engines.

An overview of theoretical problems related to 
some nano-mechanical (NMS) and nano-electro-mechanical
(NEMS) systems is given by  MacKinnon. He explicitly deals with
two models of NEMS: 1. a system of gears in which he investigates the effects
of quantization of angular momentum, and 2. a quantum shuttle.
In the discussion based on properties of these two models he shows
essential problems of NEMS models as for their  
understanding and their experimentally observable realizations: 
1. to create a model of experimentally  detectable
quantum effects related to both mechanical and 
electronic degrees of freedom, and  2. to describe properly the 
dissipation of mechanical energy.

The quantum shuttle, as a representation of NEMS, is studied
in the article of Flindt et al. They present a method for calculating
the current noise spectrum for NEMS that 
can be described by a Markovian generalized master equation.
The analysis of the gained noise spectrum shows
two possible mechanisms beyond  the current through 
the quantum shuttle device: depending on parameters,
either shuttling or sequential tunnelling will prevail.

Rekker et al investigate the classical Brownian motion of particles
under some specific constraints. They consider the noise-flatness-induced
hypersensitive transport of overdamped Brownian
particles in a tilted sawtooth potential drive by multiplicative non-equilibrium
three-level noise and additive white noise.

The following paper of Chvosta and \v Subrt is by its theme closely related
to the paper of Rekker et al. Chvosta and \v Subrt model the one 
dimensional diffusion dynamics of the Brownian
particle in piecewise linear time dependent potentials. 
They study two model potential profiles: W-shaped double well 
and a periodic array saw-tooth. In both cases, the potential
is superimposed on a  step of harmonically oscillating height.

The section ends with a study of a quantum version of  molecular 
motors. Zueco  and Garc\'ia-Palacios solve the Caldeira-Leggett master
equation in the phase-space representation to describe the behaviour
of quantum ratchets. They discuss the transition between the classical
and quantum behaviour of ratchets (in terms of methods using
Fokker-Planck as a classical version of the Langevin equation 
and Caldeira-Legget as a quantum version of the 
quantum Langevin equation) and the related decoherence processes.

\section*{Summary}
The FQMT'04 conference and the conference contributions to these
proceedings have demonstrated many relations between such areas
as quantum thermodynamics, statistical physics, 
quantum measurement theory, decoherence theory, 
physics of small systems, molecular motors and 
nano-electro-mechanical systems. Apparently,
there is also an increasing tendency for merging
theoretical and experimental methods of 
quantum optics and solid state physics. Lectures,
contributions and discussions during the conference 
have also shown several really challenging goals of the
recent physics, which are common to all these areas:\\
{\bf 1. To improve methods for the description 
of (open) systems far from equilibrium:}
We need to develop non-equilibrium theory which
will be  able to describe (open) systems with 
various numbers of particles (e.g. from individual
electron systems up to many-electron systems)
with sufficient accuracy in all time ranges,
e.g. covering processes and dynamics of the system
from short-time to long-time scales. To this end, we need to 
find a proper description of initial conditions, 
interactions in the system,   and efficient methods of  
how to find dynamics  beyond both Markovian and linear 
approximations. A really challenging problem is to develop a 
theory which describes proper dynamics of the system
when the interaction between the system and the
reservoir is a strong one,  and weak coupling
theories are not working properly. 

{\bf 2. To develop more complex models for dissipation
processes:}
In  "small systems", such as NEMS,  complicated
couplings can be created between various parts of the
system and their surroundings.
There is a possibility that e.g. the resonator can be 
damped via excitations of internal modes of the system.
The dissipation can also be mediated via the 
strong electron-phonon interaction when 
an adiabatic (Born-Oppenheimer) approach is not
sufficient. In other words, we have to study
dissipation mechanisms in these new systems and
to develop methods for including them in the 
dynamical description, so that these mechanisms
would be still practically treatable 
within the Generalized Master Equations (GME) 
framework.

{\bf 3. To improve our understanding of decoherence
in various (microscopic - mesoscopic - macroscopic) systems:}
There  is an increasing need  to understand:  a) the relation between
decoherence processes  and the quantum measurement 
problem, b) emergence of classical macroscopic 
world from the quantum world, and c) the physics of possible 
working qubit systems.
As to the first item, some progress was discussed at the meeting
by presenting an explicit, solvable model for a quantum measurement.
It would be interesting to see more research along these lines.

{\bf 4. To create new methods to analyze noise spectra
and to thereby extract useful information for systems 
such as nano-electro-mechanical systems (NEMS):} 
There is also an increasing need to gain more information
about "mesoscopic" systems from transport studies 
as opposed to only the mean current, which  measures the total charge 
transported via the system. The full counting statistics (FCS),
i.e. the knowledge of the whole distribution of transmitted
charge through the small system, of course,  provides more information
about the system than just only the first cumulant of the FCS
(mean current). Already the second cumulant, the current noise,
can help us to distinguish between the different transport 
mechanisms which lead to the same mean current.
The problem, however, is  how to coordinate 
the choice of a model of the measured small system
with a method of how to calculate  reliable several first 
cumulants; calculations  heavily depend 
on an approximation of a Generalized Master 
Equation (GME). Due to technical difficulties
calculations are up to now limited, more or less,  
to Markovian  approximations of GME of used models.   

{\bf 5. To study intensively physical processes 
in "small" biological systems, i.e. on the level of cells
and their organelles:}
Recent nano-technologies enable us to construct (biomimetic) 
systems, which mimic at least some features of complicated 
biological systems and mechanisms in living cells. 
Apart from investigation of mimetic systems,
nano-devices (e.g. NEMS) provide us a possibility to
"follow individual molecules" in cells and manipulate them. 
This increases a possibility of "symbiosis" between biology
and physics:  We can improve our knowledge of how 
cells work using physics, but also physics research 
can be motivated by studies of cellular mechanisms.
Molecular motors is the field where 
physics and biology already mutually cooperate. 
It is assumed nowadays, that every directed motion 
in living cells (such as transport of ions through cells'
membranes, and kinesin walking along
cytoskeletal filaments)  is governed by molecular 
motors. These "microscopic engines" probably 
operate in the overdamped Brownian motion regime
and for a better understanding 
of their roles in cells,  a further 
development of methods of statistical 
physics is essential: we do not deal only
with individual motors in cells but our challenge
is to understand highly cooperative behaviour
of many molecular motors, filaments of the cytoskeleton
system, transport through membranes, and organelles
of the cell. We can encounter such phenomena
as traffic flows, traffic jams and pattern formation
in cells. In fact, there are many problems
where physics can help biology and vice versa.
For example, recent investigations  show that
statistical physics can help us
understand biological 
information processing: the effect of stochastic
resonance can explain how weak biological signals
are amplified by random fluctuations.

{\bf 6. To further improve systems which we can 
study experimentally, to suggest new experiments
for small systems and to investigate various
combinations of systems and parameters we have under
our control:}
There are many promising areas of research, such as
Quantum Brownian motion and Molecular motors,
Opto-mechanical and Nano-electro-mechanical systems, 
Quantum optics and  Physics of quantum computing, 
which provide us a possibility to test experimentally the developed
models and basic  theories (as  for example the 
theory of decoherence) in greater detail. 

There is hope that working on the above mentioned
problems we will in future understand how and when a 
possible quantum thermodynamic description will appear as a 
special limit of quantum statistical physics.
We will have  better explanation for the irreversibility 
not only from the point of view of how it appears in the macroscopic
world when a microscopic description is in principle
based on a reversible description, but also in relation to
quantum measurement process which is an irreversible
process itself. At the same time, we will understand better
when and how the classical macroscopic world which
we daily observe is emerging from our quantum
statistical picture of the microworld.

Even small experimental systems (generally far from equilibrium 
states) are still complicated from the point of view of theoretical 
description  and the interpretation of experiments. 
The task to understand the phenomena discussed at the FQMT'04 conference
is to navigate in between {\it Scylla} and {\it Charibda}, the
opposing rocks,  which are created on one  side by theoretical models  
and on the other side by experiments. We need to develop 
theoretical methods and models  we are able to solve and 
from which  it is possible to  extract information comparable 
with experimental data.  At the same time,  the model has  to be 
able  to describe the  real complexity of the experiment.

To conclude, we can say that the depth and the diversity
of the questions addressed at the FQMT'04 conference were
very profound and this is reflected in these proceedings.

\section*{Acknowledgements}
We would like to thank all participants of the FQMT'04 for creating
a very good atmosphere during the conference and for their 
participation in very lively scientific discussions.
We also really appreciated all the excellent lectures
we heard during the FQMT'04 conference and we would
like to thank all lecturers for them. \\

This research was partially supported by the Grant Agencies of the
Czech Republic  and of the Academy of Sciences within the grant projects  202/04/0585
and A1010404, respectively. The research was carried out by V~.\v{S}pi\v{c}ka  within 
the Institutional Research Plan AV0Z10100521 financed
by the Academy of Sciences of the Czech Republic.

\end{document}